\documentclass[aps,prc,showpacs,twocolumn,floatfix,nosuperscriptaddress]{revtex4-1}
\usepackage{amssymb}
\usepackage{epsfig,graphics,color}
\usepackage{graphicx}% Include figure files
\usepackage{dcolumn}% Align table columns on decimal point
\usepackage{bm}% bold math
\usepackage{amsmath,eufrak,mathrsfs}
\usepackage{xcolor}

\newcommand \tie {{\it i.e.}}
\newcommand \ie {{\it i.e.} }

\newcommand \kd  {\delta}
\newcommand \ra  {\rightarrow}

\newcommand \g {\gamma}

\newcommand \x {\cdot}

\newcommand \A {\alpha}

\newcommand \lc {\langle}
\newcommand \rc {\rangle}

\newcommand \bvec{\left( \begin{array}{c} }
\newcommand \evec{\end{array} \right)}

\newcommand \eg {{\it e.g.}}
\newcommand \bea{\begin{eqnarray} }
\newcommand \eea{\end{eqnarray} } 
\newcommand \nn {\nonumber}
\newcommand {\be} {\begin{equation}}
\newcommand {\ee} {\end{equation}}
\newcommand {\epem} {$e^+ e^-$}
\newcommand {\mbx} {\mbox{}}

\newcommand {\ata} {& \times &}

\begin{document}

\title{The energy and scale dependence of $\hat{q}$ and the JET puzzle}

\author{Amit~Kumar}
\affiliation{Department of Physics and Astronomy, Wayne State University, Detroit, MI 48201.}

\author{Abhijit~Majumder} 
\affiliation{Department of Physics and Astronomy, Wayne State University, Detroit, MI 48201.}

\author{Chun~Shen}
\affiliation{Department of Physics and Astronomy, Wayne State University, Detroit, MI 48201.}

\date{\today}

\begin{abstract} 
 We present an attempt to probe the underlying structure of the quark-gluon plasma (QGP) at high resolution, based on the extracted jet transport coefficient $\hat{q}$. We argue that the exchanged momentum $k$ between the hard parton and the medium varies over a range of scales, and for $k\geq1$~GeV, $\hat{q}$ can be expressed in terms of a parton distribution function (PDF). Because the mass of a QGP constituent is unknown, we define a scaling variable $x_N$ to represent the ratio of the parton momentum to the momentum of a self-contained section of the plasma which has a mass of 1 GeV. This scaling variable is used to parametrize the QGP-PDF.  
Calculations, based on this reconstructed $\hat{q}$ are compared to data sensitive to the hardcore of jets \tie, the single hadron suppression in terms of the nuclear modification factor $R_{AA}$ and the azimuthal anisotropy parameter $v_{2}$, as a function of transverse momentum $p_{\mathrm{T}}$, centrality and energy of the collision. 
It is demonstrated that the scale evolution of the QGP-PDF is responsible for the reduction in the normalization of $\hat{q}$ between fits to Relativistic Heavy-Ion Collider (RHIC) and Large Hadron Collider (LHC) data; a puzzle, first discovered by the JET collaboration.  
\end{abstract}

\maketitle

 \section{Introduction.}
The measurement of high $p_{\mathrm{T}}$ hadron observables in high energy heavy-ion collisions is a primary tool for understanding the properties of the Quark-Gluon plasma (QGP). Such calculations require the factorization~\cite{Collins:1985ue,Collins:1988ig,Collins:1989gx} of the parton-parton hard scattering cross section from the final state fragmentation function $[D(z,Q^2)]$, the initial state Parton Distribution Functions (PDFs) $[G(x,Q^2)]$, and in-medium jet transport coefficients such as $\hat{q}$ (the transverse momentum diffusion coefficient)~\cite{Baier:2002tc} and $\hat{e}$ (the longitudinal momentum drag coefficient)~\cite{Majumder:2008zg}. The fragmentation function and PDFs are scale-dependent universal functions, extracted from experiments such as \epem~annihilation $[D(z,Q^2)]$ and Deep-Inelastic scattering data $[G(x,Q^2)]$. 
The jet transport coefficients such as $\hat{q}$ are also universal functions and set using one or two data points from leading hadron suppression in heavy-ion collisions. The factorization theorem is well established for proton-proton collision due to work by Collins, Soper, Sterman ~\cite{Collins:1985ue,Collins:1988ig,Collins:1989gx}, and generally assumed to hold for the calculation of observables at high $p_{\mathrm{T}}$ ($>$8 GeV) in heavy-ion collisions, due to the large separation of scales between the medium and the final state high 
$p_{\mathrm{T}}$ hadron. 

In the last decade, several efforts have been made to calculate the energy loss of a leading parton in the medium, in order to describe the suppression of the yield of the leading hadrons~\cite{Wiedemann:2000tf,Wang:2001ifa,Gyulassy:2000er,Arnold:2002ja,Majumder:2010qh}. These formalisms encode the modification of the hard parton by the non-perturbative transport coefficient $\hat{q}$. The first rigorous extraction of $\hat{q}$ was carried out by the JET collaboration~\cite{Burke:2013yra}, where a systematic model-to-data comparison was performed by constraining the nuclear modification factor $R_{AA}$ for all these differing energy loss formalisms.
The nuclear modification factor $R_{AA}$ measures the  suppression of the leading hadron, and is expressed as the ratio of the differential yield of hadrons $d^{2} N_{AA} ( b_{min}, b_{max} ) $ in bins of $p_{\mathrm{T}}$ , rapidity ($y$) and
centrality (codified by a range of impact parameters $ b_{min}$ to $b_{max}$ ) in a nucleus-nucleus collision, to the differential yield of hadrons in a proton-proton ($pp$)
collision, scaled by $\langle N_{bin}  ( b_{min}, b_{max} ) \rangle$, the average number of expected nucleon-nucleon collisions in
the same centrality bin:
\bea
R_{AA} = \frac{ \frac{ d^{2} N_{AA} ( b_{min}, b_{max} )  }{d^{2}p_{\mathrm{T}} dy }   } 
{  \langle N_{bin} ( b_{min}, b_{max} ) \rangle  \frac{d^{2} N_{pp} }{ d^{2 }p_{\mathrm{T}} dy } }.
\eea

The calculations done by the JET collaboration were for the most central (0-5\%) events at RHIC and LHC collision energies. These calculations were run on identical 2+1D viscous hydro-dynamical profiles from Ref.~\cite{Shen:2014vra}. 
The striking outcome of this study was the demonstration that various energy-loss formalisms exhibit a common property: The interaction strength, defined as, 
\bea
\hat{\mathscr{Q} } (T)= \hat{q}(T) / T^3,
\eea
at the same temperature $T$, is lower at the LHC compared to that at RHIC. We refer to this odd property as the JET puzzle.

So far, a few attempts to explore the possible dependence on the free parameter $\hat{\mathscr{Q} }$ been made, \eg, in
Ref.~\cite{Xu:2015bbz,Andres:2016iys}. The work of the authors of Ref.~\cite{Xu:2015bbz}, based on the possibility of magnetic monopoles in the  plasma, suggests that the interaction strength $\hat{\mathscr{Q} }$ has a non-trivial (upward cusp-like) temperature dependence in the region around $T_c$ (based on the quasi-particle relation derived in Ref.~\cite{Majumder:2007zh}). This implies that experiments at RHIC are more sensitive to this rise in $\hat{\mathscr{Q}}(T)$ near $T_c$, due to lower initial temperatures at RHIC, compared to the LHC. As a result, the effective $\hat{\mathscr{Q}}$ extracted in comparison with data tends to be higher at RHIC than at LHC. 
However, studies done in recent years by the authors of Ref.~\cite{Andres:2016iys}, are in clear contradiction with such a prediction. Moreover, the work of the authors of Ref.~\cite{Andres:2016iys}, based on the ASW formalism reveals that the interaction strength $\hat{\mathscr{Q} }$ is sensitive to the center-of-mass energy of nucleus-nucleus collision rather than the local temperature of the QGP or the centrality of colliding nuclei.

In this paper, we propose an alternative formalism to study the transport coefficient $\hat{q}$ and demonstrate that at both RHIC and LHC, both the centrality dependence of the $R_{AA}$ and 
azimuthal anisotropy ($v_2$) of leading hadrons can be well described using a $\hat{\mathscr{Q}}(T)$, 
that has no such cusp-like behavior near $T_c$ (in this effort, similar to Ref.~\cite{Burke:2013yra}, we will ignore the minor effect of $\hat{e}$~\cite{Qin:2012fua,Qin:2014mya} on light flavors).
Note that the initial temperature achieved in a heavy-ion collision (at thermalization) not only depends on the center-of-mass energy of nucleus-nucleus collisions but also on its centrality.
The effect of the cusp at $T_{c}$ as proposed in Ref.~\cite{Xu:2015bbz} should be much stronger in peripheral collisions compared to most central collisions, leading to noticeably larger suppression than expected based on a monotonic scaling relation between $\hat{q}$ and $T$. However, no such effect has been found in the current work. This is also consistent with recent observations within the ASW formalism as reported in Ref.~\cite{Andres:2016iys}. 

In this article, we formulate a transport coefficient $\hat{q}$ that depends not only on the local temperature $T^3$ but also the energy of the leading parton and the scale at which the leading parton probes the medium. With this, we successfully demonstrate the reduction of $\hat{\mathscr{Q} }$ at LHC compared to RHIC.
Preliminary work has been reported in Ref.~\cite{KUMAR2017536,Bianchi:2017wpt}. Following similar ideas, the reader may also find a study reported in 
Ref.~\cite{PengRuXing2019}, where the authors used a similar form of $\hat{q}$ to carry out a global analysis for cold nuclear matter.

 Based on estimates from most hydrodynamical simulations of QGP, we note that the highest initial temperature achieved at the LHC collision energy is a bare 20\% higher than at RHIC collision energy. But, the energy of the leading parton at the LHC (hadron $p_{\mathrm{T}}\sim$100~GeV) is an order of magnitude higher than those of RHIC (hadron $p_{\mathrm{T}}\sim$10~GeV). Therefore, the leading partons have a considerably (though logarithmically) higher virtuality at the LHC, a smaller transverse size. At these short distance scales, a section of QGP probed by the leading parton may appear more dilute. This effect is similar to the Dokshitzer-Gribov-Lipatov-Altarelli-Parisi (DGLAP) evolution of PDFs at finite and large $x$, with increasing $Q^{2}$~\cite{Gribov:1972ri,Gribov:1972rt,Altarelli:1977zs,Dokshitzer:1977sg}.  

The remaining paper is organized as follows: In Sect.~\ref{setup}, we outline the basic assumptions and highlight the salient formulas of the single scattering induced emission within the higher twist energy loss scheme. This section points out how coherence like effects may be included within the higher-twist scheme.  In Sect.~\ref{DGLAP}, we extend the single scattering induced emission to a multiple emission formalism and motivate the medium modified DGLAP evolution equation. This section highlights the assumptions made and the dependencies of the medium modified fragmentation function. In Sect.~\ref{Sect:QGP}, we outline a model for $\hat{q}$ based on the QGP-PDF. This in combination with the scale of the hard parton demonstrates how the resolution scale of the jet increases with its virtuality. 
In Sect.~\ref{raa_v2}, we present numerical calculations based on our model and compare to data. Comparison with the $R_{AA}$ and $v_2$ across two different collision energies and four centralities are carried out. The probable form of the QGP-PDF is extracted. Concluding discussions are presented in Sect.~\ref{summary}.

\section{Setup and Simple Estimates.}
\label{setup}
In this section, we discuss the effect of scale evolution on the resolution of the medium.
Let us consider an analogous process of Deep-Inelastic Scattering on a large nucleus ($A \gg 1$) 
at large photon virtuality ($Q^{2} \gg \Lambda_{QCD}^{2}$) and 
focus on the limit where a hard quark is produced. 
In this limit, we can factorize the propagation of the quark 
in the extended medium from the production process, 
obtaining equations for the scattering induced single gluon emission spectrum or the transverse momentum distribution of the produced quark. 
We carried out our analytical calculations in the Breit frame, where
the virtual photon $\g$ and the nucleus have 
momentum four-vectors $q =\left[ -Q^2/2q^-, q^-, 0, 0\right]$ and $  
P_A \equiv A[p^+,0,0,0]$, respectively, where $A$ is the mass number and $q^-$ is the large light-cone momentum of the hard quark.  We choose the nucleus to be traveling in the positive $z$ direction with large light-cone momentum $P^{+}=Ap^{+}$ and the photon in the negative $z$ direction. In this frame, the Bjorken variable is given as $x_B = Q^2/2p^+q^-$. 
We define the momentum of a struck quark or gluon in any of the 
nucleons in terms of momentum fraction $x$ as $p_{q,g} \simeq xp$ (where $0<x<1$). In this picture, each nucleon is time dilated to an almost static state with an arbitrary number of near on-shell quarks and gluons, which are described by the Parton Distribution Function (PDF) of the nucleon (in this paper we will ignore nuclear effects on the nucleon PDF).

{Now, we consider a scenario where a hard quark propagates through the nucleus. It undergoes multiple scatterings off the gluon field of the nucleus and radiates gluons.
Without any scattering or radiation, the transverse momentum ($\vec{k}_{\perp}$) distribution of the quark is given approximately by a $\kd$-function, $\frac{d \sigma}{\sigma d^2k_\perp} \propto \delta^{(2)}(\vec{k}_{\perp})$.
Scattered by one medium parton, illustrated in Fig.~\ref{qhat}, the hard quark's averaged transverse momentum is shifted by $\int d Y^{-} \hat{q} (Y^-)$. 
The effect from further uncorrelated multiple scatterings on hard quark's transverse momentum distribution can be modeled as a diffusion process described in Ref.~\cite{Majumder:2007hx}.
Using either the single scattering equation Ref.~\cite{Guo:2000nz,Majumder:2004pt,Fries:2000da} or the full diffusion equation in Ref.~\cite{Majumder:2007hx} one obtains the fundamental relation that defines the transport coefficient,
\bea
\lc k_\perp^2 \rc = \!\!\!\int \!\! d^2 k_\perp  k_\perp^2 \frac{d\sigma}{\sigma d^2 k_\perp} =  \!\!\!\int_0^{L^-} \!\!\!\!\!\!\!\!dY^- \hat{q} (Y^-) \simeq \hat{q} L^-. \label{qhatL}
\eea
The last equality in the equation above is valid in the limit that $\hat{q}$ is constant along the path traversed by the hard quark.
\begin{figure}[htbp]
\resizebox{2.9in}{1.6in}{\includegraphics{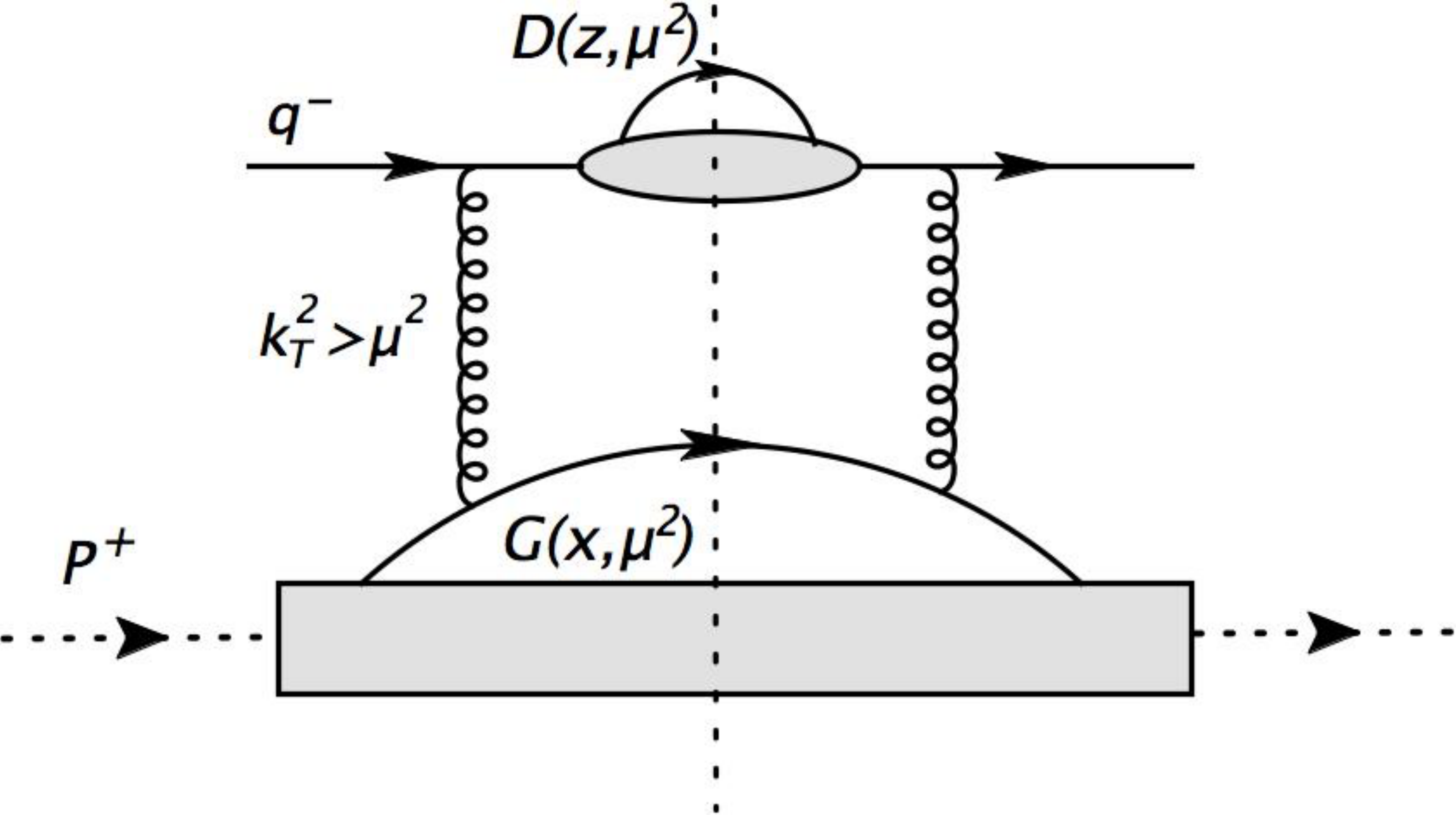}} 
    \caption{A forward scattering diagram at leading order (LO).  Here, a hard quark with large light-cone momentum $q^-$ scatters via a Glauber gluon $k_\perp $with a hard collinear fluctuation in a target constituent (nucleon in nucleus or QGP constituent).  The vertical dotted line represents the cut-line.}
    \label{qhat}
\end{figure}

The derivation of Eq.~\eqref{qhatL} in the case of multiple scattering rests on the assumption that the scatterings are independent of each other. In the case of DIS on a large nucleus, this implies that subsequent scatterings take place on separate nucleons, or that double scattering on a single nucleon is rare. This assumption holds for the case of high energy partons. 
In this limit, scattering also induces radiation, in addition to the vacuum radiation emanating from the quark. For the case where only the radiated gluon scatters, the differential yield of induced gluons, in bins of light-cone momentum fraction $y$ and transverse momentum $\vec{l}_\perp$, from a single scattering can be expressed as~\cite{Wang:2001ifa,Zhang:2003yn,Aurenche:2008mq}, 
\bea
\frac{dN_g}{dyd l_\perp^2} &=& \frac{\A_s}{2\pi^2}  P(y)\!\! 
\int \!\!\frac{d \zeta^-  d \kd \zeta^- d^2 k_\perp d^2 \zeta_\perp}{(2\pi)^2} \label{gluon_scattering} \\
\ata \frac{ 2 - 2 \cos\left\{  \frac{ ( l_\perp - k_\perp )^2 \zeta^- }{2 q^- y (1-y)}   \right\} }{ ( l_\perp - k_\perp )^4 } 
e^{ -i \frac{k_{\perp}^{2}}{2 q^{-}} \delta \zeta^{-} + i \vec{k}_{\perp} \x  \vec{\zeta}_{\perp} } \nn \\
\ata  \left\lc p_{B} \left|  {A^a}^+ (\zeta^{-} \!\!\!+ \kd \zeta^{-}, \vec{\zeta}_{\perp})    {A^{a}}^{+}_{\alpha} (\zeta^{-},0_\perp) \right| p_{B} \right\rc . \nn
\eea
In the equation above (and the remainder of this paper), we have neglected the obvious initial state and hard-scattering factor $G(x,Q^2) \hat{\sigma}$, we assume that the photon light-cone momentum $q^-$ and the $Q^2$ are measured in some restricted  range, then the Bjorken-$x$ is determined such that $x = Q^2/(\sqrt{2}Mq^-)$.
In the equation above, $P(y)$ is the regular AP splitting function.
The scattering takes place at the location $(\zeta^-,0_\perp)$ in the amplitude and at the 
shifted location $(\zeta^- + \delta \zeta^-, \vec{\zeta}_\perp)$ in the complex conjugate, where $A^{a+}$ represents the dominant component of the gluon vector potential at these locations (in $A^{a-} = 0$ gauge), in the nucleon state $| p_B \rc$.

Requiring that the final nuclear state be identical in both amplitude and complex conjugate, the $\kd \zeta^-$ and $\zeta_\perp$ integrals are limited by the size of a nucleon, i.e., the same nucleon that is struck in the amplitude, is also struck in the complex conjugate. The $\zeta^-$ integral is limited by the formation time of the radiation $\tau^-$ or the light-cone length $L^-$ that has to be traversed in the nucleus by the parton.
The scattering exchanges a transverse momentum $\vec{k}_\perp$ between the radiated gluon and the medium, along with ($+$)-momentum $k^2_\perp/(2q^-)$, and ($-$)-momentum $k^2_\perp/(2p^+) \ll q^-$. The ($-$)-momentum exchanged has already been integrated out in Eq.~\eqref{gluon_scattering}, as it makes a negligible change in the ($-$)-momenta of the produced quark and gluon. 

\begin{figure}[htbp]
\resizebox{2.75in}{2.5in}{\includegraphics{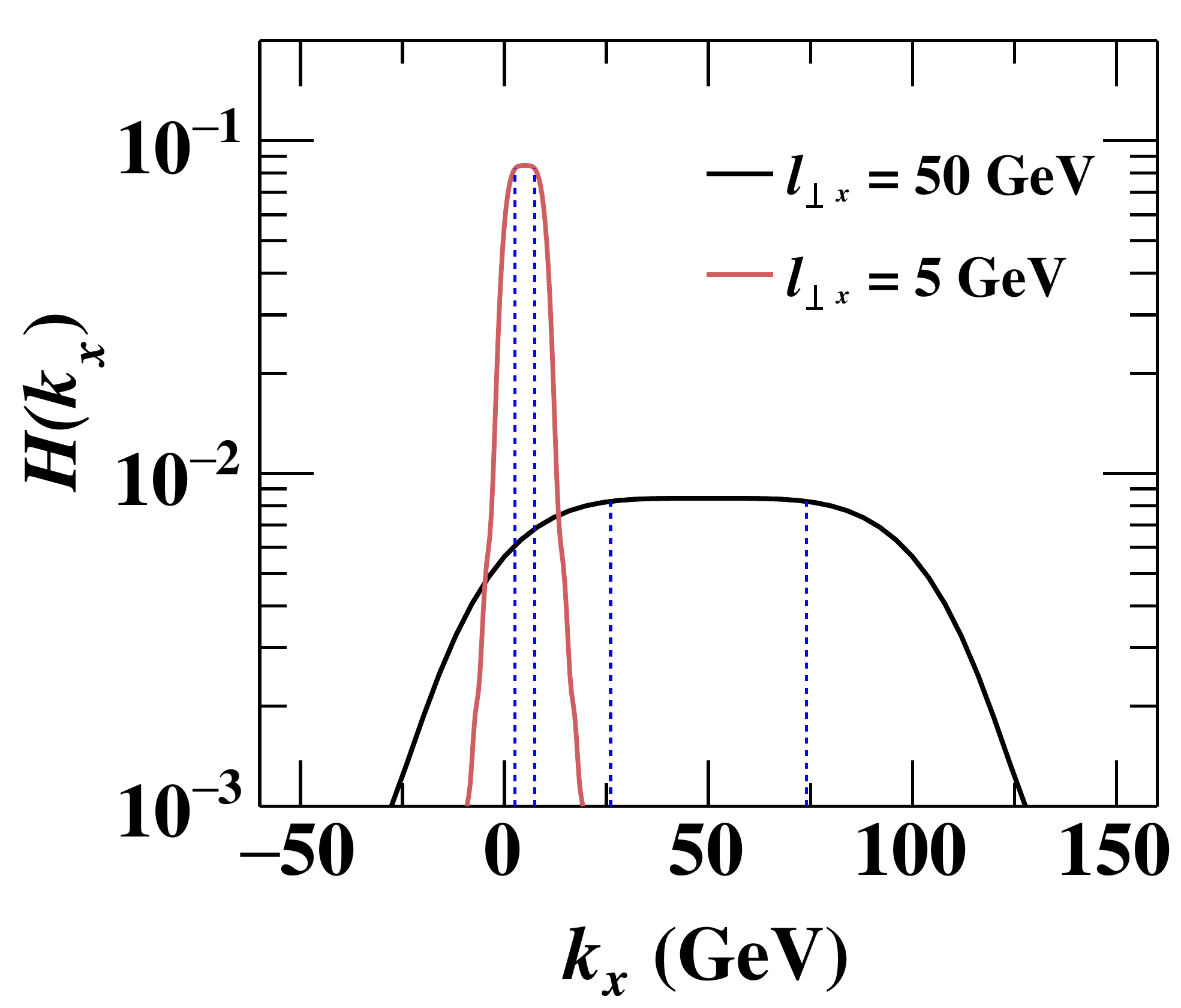}} 
%\vspace{0.25cm}
    \caption{Plot of Eq.~(\ref{eq7})  for $l_y = k_y = 0$ GeV, $q^{-}=50$ GeV, and $y = 0.5$, as function of $k_x$ for two choices of $l_x$. The blue dashed lines at $k_x \sim l_x \pm (l_x/2)$ isolate the portion of the function where the value is within 95\% of the maximum value.}
    \label{plot_integral}
\end{figure}

While not made clear in prior publications, there is a tension between the 2nd and 3rd lines of Eq.~\eqref{gluon_scattering}, regarding the dominant range of the 
$k_\perp$ integration. While explicitly demonstrated in the subsequent sections, it is immediately clear that the expectation of the operator product in the 3rd line of Eq.~\eqref{gluon_scattering}, is enhanced for separations $\zeta_\perp \sim 1/{p_{B}}_\perp \sim 1/\Lambda_{QCD}$. 
Thus, the matrix element prefers $k_\perp \sim \Lambda_{QCD}$. 
However, the 2nd line of Eq.~\eqref{gluon_scattering} prefers $k_\perp \sim l_\perp$, the transverse momentum of the radiated gluon. 
This is consistent with the idea of \emph{coherence} first expounded in 
Refs.~\cite{Armesto:2011ir,MehtarTani:2011tz,CasalderreySolana:2012ef}. 
To clearly illustrate this, we briefly make the approximation that, in a large nucleus, one can assume longitudinal translation invariance of the matrix element, i.e., on average, all nucleons are similar and thus, 
\bea
&& \left\lc p_{B} \left|  {A^a}^+ (\zeta^{-} \!\!\!+ \kd \zeta^{-}\!\!, \vec{\zeta}_{\perp})    {A^{a}}^{+}_{\alpha} (\zeta^{-}\!\!) \right| p_{B} \right\rc \\
&\simeq& \left\lc p_{A} \left|  {A^a}^+ ( \kd \zeta^{-}\!\!, \vec{\zeta}_{\perp})    {A^{a}}^{+}_{\alpha} (0) \right| p_{A} \right\rc . \nn
\eea
In the above expression, nucleon $A$ is at the origin $\vec{\zeta}=0$, and nucleon $B$ is at ($\zeta^-,0$). 
In this approximation, we can factorize the location integrals into the second and third lines of Eq.~\eqref{gluon_scattering}, obtaining
\bea
\frac{dN_g}{dyd l_\perp^2} &=&  \frac{\A_s}{2\pi^2}  P(y)
\int \frac{  d^2 k_\perp}{(2\pi)^2} \label{gluon_scattering_factorized} \\
\ata \int_0^{\tau^-} d \zeta^-  \frac{ 2 - 2 \cos\left\{  \frac{ ( l_\perp - k_\perp )^2 \zeta^- }{2 q^- y (1-y)}   \right\} }{ ( l_\perp - k_\perp )^4 } \nn \\
\ata \int d \kd \zeta^-  d^2 \zeta_\perp e^{ -i \frac{k_{\perp}^{2}}{2 q^{-}} \delta \zeta^{-} + i \vec{k}_{\perp} \x  \vec{\zeta}_{\perp} } \nn \\
\ata \left\lc p_{B} \left|  {A^a}^+ ( \kd \zeta^{-}, \vec{\zeta}_{\perp})    {A^{a}}^{+}_{\alpha} (0) \right| p_{B} \right\rc . \nn
\eea

We define the integral in the second line of the Eq. (\ref{gluon_scattering_factorized}) as
\bea
H(k_\perp, l_\perp, q^{-}, y) = \int_0^{\tau^-} d \zeta^- \frac{ 2 - 2 \cos\left\{  \frac{ ( l_\perp - k_\perp )^2 \zeta^- }{2 q^- y (1-y)}   \right\} }{ ( l_\perp - k_\perp )^4 },  \label{sampling_integral}
\label{eq7}
\eea
where the formation time $\tau^{-} = 2q^-/\mu^2$ and $\mu^2 \sim l_\perp^2$.
This function (normalized) is plotted in Fig.~\ref{plot_integral} as a function of $k_{\perp,x}$ for two different values of ${l_\perp}_x $= 5,  50 GeV. We choose the projection along $k_{\perp,y} = l_{\perp,y}=0$, $q^{-}=50$ GeV, and $y = 0.5$.
 As one immediately notes, increasing $l_{\perp}$, which in transverse space corresponds to making the quark-gluon dipole smaller, selects $k_{\perp}$ exchanges that are of the same order as $l_{\perp}$, i.e., wavelengths that can resolve the quark-gluon dipole as separate partons. 

To completely uncouple the terms in the second line with those in the third and fourth line of Eq.~\eqref{gluon_scattering_factorized}, a Taylor expansion in $k_\perp^2/l_\perp^2$ (odd powers vanish due to cylindrical symmetry) is carried out. Using by parts integration, the product $ \int d^{2} \zeta_{\perp} i\vec{k}_{\perp}\exp[ i \vec{k}_{\perp} \cdot \vec{\zeta}_{\perp} ]  A^{a+}( \vec{\zeta}_{\perp})$ can be converted to $\int d^{2} \zeta_{\perp} \exp[ i \vec{k}_{\perp} \cdot \vec{\zeta}_{\perp} ] \partial_{\perp} A^{a+}( \zeta_{\perp}) \simeq \int d^{2} \zeta_{\perp} \exp[ i \vec{k}_{\perp} \cdot \vec{\zeta}_{\perp} ] F^{a+}_{\perp}( \vec{\zeta}_{\perp} )$, where $F^{a+}_\perp$ is the gluon chromo-magnetic field transverse to the direction of propagation of the hard parton.  
The first non-vanishing contribution, from the term $k_{\perp}^{2}/l_\perp^2$, is contained within the transport coefficient $\hat{q}$, 
now written with the $\zeta^-$ dependence re-introduced as, 
\bea
\hat{q} (\zeta^-) &=&  \int \frac{d\kd\zeta^{-}}{2 \pi} \int \frac{d^{2} k_{\perp}  d^{2} \zeta_{\perp}}{(2 \pi)^{2}} 
e^{ -i \frac{k_{\perp}^{2}}{2 q^{-}} \zeta^{-} + i \vec{k}_{\perp} \x  \vec{\zeta}_{\perp} }  \label{qhatEqn}\\
\ata \left\lc p_{B} \left|  {F^{a}}^{+ \alpha} (\zeta^{-} \!\!\!+ \kd \zeta^{-}, \vec{\zeta}_{\perp}) {F^{a}}^{+}_{\alpha} (\zeta^{-}) \right| p_{B} \right\rc  . \nn
\eea

It should be pointed out that while coherence effects (2nd line of Eq.~\eqref{gluon_scattering_factorized} above), require a $k_\perp \sim l_\perp$ for energy loss to take place on a virtual parton, momentum fluctuations where $k_\perp \gg \Lambda_{\rm QCD}$, are suppressed in a nucleon. As a result, 
$\hat{q}\tau^- \ll l_\perp^2$. 
Extending this argument further, it will be assumed that higher power corrections \eg, $k_\perp^4/l_\perp^4$, which yield terms such as 
\bea
\mbx\!\!\!\!\!\!J\!\!=\!\!\frac{ \left\lc p_{B} \left|  {\partial^\perp F^{a}}^{+ \alpha} (\zeta^{-} \!\!\!+ \kd \zeta^{-}, \vec{\zeta}_{\perp}) {\partial_\perp F^{a}}^{+}_{\alpha} (\zeta^{-}) \right| p_{B} \right\rc }{l_\perp^4}, \label{HigherOrder}
\eea
are suppressed compared to the $\hat{q} \tau^-/l_\perp^2$, and are ignored in the remainder of this paper. The reader will note that this statement depends on assumptions regarding the distribution of momentum originating from the nucleon state. In Sect.~\ref{Sect:QGP}, we will present a model which obeys this approximation. Given these, we obtain the medium induced spectrum of gluons radiated from a single scattering in the medium as 
\bea
\frac{dN_g}{dy d^2 l_\perp} &=& \frac{\alpha_S P(y)}{2\pi^2} \int_0^{\tau^-} d \zeta^- \hat{q}(\zeta^-) \nn \\
\ata \frac{ 2 - 2 \cos\left\{  \frac{ ( l_\perp )^2 \zeta^- }{2 q^- y (1-y)}   \right\} }{ ( l_\perp )^4 }. \label{gluon_emission_with_qhat}
\eea
In the subsequent section, we use the above formula to compose a multiple emission formalism to compute the yield of hadrons fragmenting from the hard parton. }

\section{Multiple Emissions and medium modified DGLAP}
\label{DGLAP}

{
In this paper, we will focus on high momentum single hadron production. In the next section, we will extend this to nuclear collisions, at RHIC this implies a $p_T \geq 8$ GeV, while at the LHC we will restrict calculations to $p_T \geq 10$ GeV. Hard partons start out with a virtuality that is much higher than any medium scale. In this stage, medium modification is power suppressed, and is a perturbative correction to the virtuality ordered vacuum-like emissions from the hard parton. 
In this paper, it will be assumed that partons remain in this state as they exit the large nucleus (dense medium), 
and then fragment in vacuum to produce hadrons. 
This picture is obviously not true for most of the hadrons that emanate from this process, 
many will be produced within or will be affected by the nuclear medium. 
However, this picture is appropriate for the highest energy hadrons which are produced in the fragmentation of the highest energy parton, 
which is expected to exit the nuclear medium prior to fragmentation.

In the case of no stimulated emission, the yield of hadrons carrying a momentum fraction $z$ of the original parton from the fragmentation of a single parton with virtuality $\mu_0^2$, is obtained using the fragmentation function $D(z, \mu_0^2)$. The change of the yield due to multiple emissions from a parton with a different virtuality $\mu^2$ is obtained as, 
\bea
D(z,\mu^2) &=& D(z,\mu_0) \\ 
&+& \int\limits_{\mu_0^2}^{\mu^2} \frac{d \mu_1^2}{\mu_1^2} \int\limits_z^1 \frac{dy}{y} P_+ (y) D\left( \frac{z}{y} , \mu_0^2 \right) + \ldots , \nn
\eea
where the $\ldots$ represent 2 and the higher number of emissions. The subscript $+$ on the splitting function indicates that we have subtracted 
the virtual correction which contains the product of the leading amplitude, and the next-to-leading order complex conjugate for no emission (and vice-versa). This virtual correction removes the infra-red divergence from soft gluon emission in the splitting function.   
Contributions from multiple emissions can be resummed by solving the DGLAP evolution equation, 
\bea
\frac{\partial D (z, \mu^2)}{\partial \log{\mu^2}} = \frac{\A_S}{2\pi} \int \frac{dy}{y} P_+ (y) D\left( \frac{z}{y} , \mu^2 \right). \label{vac_DGLAP}
\eea

In order to add the contribution from medium induced emission, we simply convert from transverse momentum to virtuality $l_\perp^2 = \mu_1^2 y (1-y)$ in Eq.~\eqref{gluon_emission_with_qhat}, and integrate $\mu_1^2$ from $\mu_0^2$ to $\mu^2$. The ensuing term can be added to the kernel of the DGLAP equation above to obtain the medium modified DGLAP evolution equation. 
Given the form of Eq.~\eqref{gluon_emission_with_qhat}, one necessarily obtains a medium modified fragmentation function $D(z,\mu^2,\zeta_i^-)|_{q^-}$, which additionally depends on the light-cone momentum $q^-$ of the hard parton, and on the location $\zeta_i^-$, where the parton is produced in the medium ($z$ is the momentum fraction carried by the hadron with respect to the parent parton, and $\mu^2$ is the scale of the function, or virtuality of the parton). This is obtained as a solution to the medium modified DGLAP evolution equation\cite{Wang:2001ifa,Majumder:2009zu}, 
\bea
&&\mbox{}\!\!\!\!\!\!\!\! \frac{ \partial \left. \!\! D\!\!\left( z, \mu^2 , \zeta_i^- \right)\right|_{q^-} } { \partial \log \mu^2 } \! = \! \frac{ \alpha_S } { 2 \pi } \!\!
\int\limits_z^1 \!\! \frac{d y} { y } \! \left[  P_+ \!\left( y \right)\!D\!\left(\! \frac{ z } { y }, \mu^2, \zeta_i^- \! \right) \right|_{y q^-}  \nn \\
&+& \left(\frac{P(y)}{y \left( 1 - y \right)}\right)_+ 
\left. D \left( \frac{ z } { y }, \mu^2, \zeta_i^- + \tau^- \right)\right|_{yq^-} \nn \\
&\times& \left. \int\limits_{ \zeta_i^- }^{ \zeta_i^- + \tau^- } \!\!\!\!d \zeta^- \frac{ \hat{ q } \left( \zeta^- \right) } { \mu^2  }
\left\{ 2 - 2 \cos \left( \frac{ \zeta^- } { \tau^- } \right) \right\}  
 \right]. 
\eea

The first line of the equation above, is identical to the evolution equation of the vacuum fragmentation function, except that the functions now also depend on the origin of the hard parton. The second and third line represent the medium modified portions, which mix functions at location $\zeta_i^-$ with partons formed ahead in the medium by a formation time $\tau^- = 2 q^-/\mu^2$. The reader will note that we mention the parent parton's energy 
$q^-$ (or $yq^-$) separate from the other variables in the fragmentation function. This is because, the rescaling of the energy on the right-hand side of  the DGLAP equation takes place in the case of  Eq.~\eqref{vac_DGLAP} as well, but is usually suppressed as vacuum fragmentation functions are invariant under boosts in the parton's direction. In the presence of a medium, the fragmentation functions are no longer boost invariant, however, parametrically this is not a new dependence such as the position dependence.
}

\section{Calculating $\hat{q}$ and its dependence on jet scale in a QGP}
\label{Sect:QGP}

{ In the preceding sections we discussed the underlying assumptions and approximations that 
constitute the higher-twist scheme of energy loss and the formalism of medium modified fragmentation functions. 
While much of this is standard knowledge, it is important to recapitulate it here as it affects the phenomenological calculation of $\hat{q}$ in terms of a parton distribution function (PDF) in a dense medium. 

Consider a hard quark or gluon propagating through a large nucleus undergoing multiple scattering off the gluon field within the nucleons in the large nucleus. Consider this process in the Breit frame where the nucleus itself is traveling towards the hard parton. A portion of the gluon field inside boosted nucleons at a resolution scale $\mu^2 \gg \Lambda_{\rm QCD}^2$ may be described as independent partons. Thus, without any prior knowledge, we can decompose $\hat{q}$ into its perturbative (P) and non-perturbative (NP) parts:
\bea
\hat{q} = \int_0^{\mu_0^2} dk_\perp^2 \frac{d \hat{q}_{NP} }{d k_\perp^2} + \int_{\mu_0^2}^{K^2} dk_\perp^2 \frac{d\hat{q}_P}{dk_\perp^2}.
\eea
In the equation above, we identify a separation scale $\mu_0^2$. For transverse momentum ($k_\perp$) exchanges that are softer than this scale, the contributions to $\hat{q}$ cannot be computed in perturbation theory. For exchanges $k_\perp^2 > \mu_0^2$ up to the maximum allowed limit $K_\perp^2$, the contributions to $\hat{q}$ can be calculated in perturbation theory assuming a single gluon exchange between the jet parton and a parton obtained from the PDF of the medium. This is the process highlighted in Fig.~\ref{qhat}. In what follows, we assume $\mu_0^2 = 1$~GeV$^2$.

A close examination Eq. (\ref{eq7}) and Fig.~\ref{plot_integral} indicates that for a given virtuality of the hard parton ($\mu^2 \sim l^2_\perp$), the dominant integration range of $k_\perp^2$ for $\hat{q}$ lies in $\mu \pm \mu/2$.
In fact, as the virtuality (or $l_\perp^2$) increases, the range of $k_\perp^2$ that contributes to $\hat{q}$ also shifts to a higher interval. This is consistent with the idea of coherence~\cite{Armesto:2011ir,MehtarTani:2011tz,CasalderreySolana:2012ef}: Smaller dipoles are affected by gluons from the medium that have a wavelength of the order of the size of the dipole. The first outcome of this observation is that at virtualities $\mu^2 \gg \mu_0^2$, we can approximate $\hat{q} \simeq \int_{\mu_0^2}^{K^2} dk_\perp^2 d\hat{q}_P/ dk_\perp^2$, i.e., $\hat{q}$ can be approximated by the perturbative contribution only. 

To further mimic this effect we define the effective $\hat{q}(\mu^2)$ as a function of virtuality $\mu^2$ of the hard jet parton. A simple choice is to integrate the range between the blue dashed lines in Fig.~\ref{plot_integral}, i.e., 
\bea
\hat{q} (\mu^2) = \int_{(\mu-  \Delta \mu)^2}^{(\mu + \Delta \mu)^2} dk_\perp^2 \frac{d \hat{q}_P}{dk_\perp^2}. 
\label{first_qhat_eqn}
\eea 
The problem with the formula above is that the integral ranges from $\mu^2/4$ to $9\mu^2/4$, i.e., $ \Delta \mu \sim \mu/2$; this will involve carrying out a perturbative calculation for $\hat{q}$ in a region where it is decidedly non-perturbative. In realistic 
calculations of the single hadron $R_{AA}$ and $v_2$, $\mu^2$ ranges from $p_T^2$ down to $\mu_0^2 = 1$ GeV$^2$.

We deal with this issue in two ways: In method $[A]$, we keep the range of the integral from $\mu^2/4$ to $9\mu^2/4$, but restrict the scale of all quantities within the integrand to be larger than $\mu_0^2$:
\bea
\hat{q}^{[A]} (\mu^2) = \int_{\mu^2/4}^{9\mu^2/4} dk_\perp^2 \frac{d \hat{q}_P}{dk_\perp^2} [{\rm Max}(k_\perp^2, \mu_0^2)]. \label{A}
\eea

We will also use an alternate formula $[B]$ which strongly restricts the lower bound of the integral, but ensures the integrand is always evaluated at 
$k_\perp^2$,
\bea
\hat{q}^{[B]} (\mu^2) = \int_{\mu^2}^{9\mu^2} dk_\perp^2 \frac{d \hat{q}_P}{dk_\perp^2} [k_\perp^2]. \label{B}
\eea
One could argue that we have simply redefined $\mu^2$ to represent the lower bound of the integral, rather than the midpoint. 
This is a minor correction, as long as the upper and lower limits of the integral are of the same order, which is the case here. Alternatively, we could argue that we have redefined the meaning of $\hat{q} (\mu^2)$ to be the inclusion of interactions from $\mu^2$ to $9 \mu^2$ (\ie inclusion of interactions at the scale of $\mu^2$).  

A more thorough effort would involve using the midpoint of the range of $k_\perp^2$ to define $\hat{q}$, as in Eq.~\eqref{first_qhat_eqn}.
However, this would involve introducing a non-perturbative $\hat{q}$ for the portion $k_\perp^2 < \mu_0^2$, which in this analysis would amount to the introduction of another tunable parameter. As the goal of the current paper is to clearly demonstrate that the rising virtuality of hard jet partons, at higher energies, leads to the sampling of a diminishing $\hat{q}$ at a higher resolution scale, we ignore the extra non-perturbative contribution and focus solely on the perturbative part by defining $\hat{q}$ by using prescription $[A]$ or $[B]$. This is an approximation that will be improved upon in a future effort.

Considering the origin of the perturbative $\hat{q}$ to be from a single gluon exchange with a parton obtained from the Parton Distribution Function (PDF) of the target, we obtain the expression for the perturbative $\hat{q}$ as a product of the density of nucleons in the nucleus at a light-cone location $\zeta^-$, the PDF of that nucleon, and the $2\ra 2$ hard scattering cross section, weighted by the square of the transverse momentum exchanged, 
\bea
\mbox{}\!\!\!\!\!\hat{q}(\mu^2\!, q^-\!\!, \zeta^-)\! 
=\!\!\!\! \int \!\!\!d^2 k_\perp  \frac{k_\perp^2 d^2 \hat{\sigma}}{d^2 k_\perp} \rho(\zeta^-) 
 \!\!\!\!\!\!\!\int\limits_{\frac{ k^2_\perp }{2p^+ q^-}}^1 \!\!\!\!\!dx G (x, k_\perp^2).
\eea
Instead of using the local $\hat{q}$, we average over an appropriate length traversed by the hard parton prior to emission denoted as $L^-$. 
This is typically the emission length, equal to the formation length of the radiation, set by the off-shellness and the light-cone momentum of the hard parton (when the formation length exceeds the length of the medium, $L^-$ represents the medium's length), 
\bea
\hat{q}(\mu^2, q^-) &=& \frac{1}{L^-} \int d^2 k_\perp  \frac{k_\perp^2 d^2 \hat{\sigma}}{d^2 k_\perp} 
 \int\limits_{\zeta_i^-}^{\zeta_i^- + L^-} d \zeta^- \rho(\zeta^-) \nn \\
\ata \int_{\frac{ k^2_\perp }{2p^+ q^-}}^1 \!\!\!\!\!dx G (x, k_\perp^2).  \label{qhat_mu_9mu}
\eea

In both equations above, $x$ refers to the momentum fraction of a parton within a nucleon, which scatters via single gluon exchange with the jet parton. The jet parton is formed in a prior hard interaction at the location $\zeta_i^-$. It interacts with a transverse gluon from a nucleon within a light-cone distance $L^-$.
In the equation above, we have not mentioned the limits of integration of $k_\perp^2$, this depends on the prescription used [prescription $[A]$ or $[B]$ as described in Eqs.(\ref{A},\ref{B})]. Given the range of integration of the $k_\perp^2$ integral, we can also replace $G(x,k_\perp^2) \sim G(x,\mu^2)$. This replacement has almost a minimal effect on the calculation of $\hat{q}$ using Eq.~\eqref{qhat_mu_9mu}.

Extending the above formalism to jets traversing through a QGP depends strongly on the assumption of independent scatterings, valid for high 
energy partons. In all calculations of perturbative QCD based jet quenching in a QGP, one assumes subsequent scatterings to be independent of each other, thus all calculations 
obtain Eq.~\eqref{qhatL}, and as an extension that the energy lost by a hard parton $\Delta E \propto \hat{q} L^2$. In a large nucleus, one can identify the nucleon radius as the 
maximal distance over which the gluon fields may be correlated. In the case of a QGP, this distance is the much smaller quantity of the Debye screening length: 
Locations beyond this 
length are not correlated. 
The Debye length is, however, temperature dependent, and as such a hard parton traverses several different screening lengths on its way 
out of the QGP.

The second extension relates to the definition of the momentum fraction. We define a ``nucleonic'' $x_N = \hat{q} L/ (2 M_N E) $, where $M_N$ is the mass of a nucleon ($=1$ GeV in the following), 
and $E$ is the energy of the jet in the rest frame of the struck portion of the QGP. 
We define this quantity due to a lack of knowledge of $m_{QGP}$, the mass of a degree of freedom, or a correlated enclosure within a QGP; 
all estimates of this mass place it at $m_{QGP} \sim gT \lesssim M_{N}$. Thus at best, our $x_N = \xi (T) x$, 
where $x$ is the actual momentum fraction of a parton within a QGP degree of freedom, and $0<\xi(T)<1$ is a temperature dependent scaling factor.

For intermediate values of $x_N$, the scale and energy dependent $\hat{q}$ is obtained from the diagram in Fig.~\ref{qhat}: Partonic 
fluctuations (indicated by solid lines with arrows), which may be a quark or a gluon at a scale $\sim \mu^2$, collinear with the target state of a nucleon or a QGP constituent (with target momentum $p^+$), radiate 
a Glauber gluon~\cite{Idilbi:2008vm}, with resolved transverse momentum $k_\perp^2 \sim \mu^2$, which scatters off the outgoing jet parton, with momentum $q^-$, 
\bea
\hat{q}(\mu^2, q^-) &=& \frac{1}{L^-}\int d^2 k_\perp k_\perp^2  \frac{ d^2 \hat{\sigma}}{d^2 k_\perp} 
\int_0^{L^-} \!\!\!dy^- \rho(y^-) \nn \\
\ata \int_{\frac{ k^2_\perp }{2p^+ q^-}}^1 dx_N G (x_N, \mu^2). \label{qhatExpression}
\eea
In the equation above, $G (x_N, \mu^2)$ represents the parton distribution at the scale $\mu^2$ of a degree of freedom in the target (the density of which is $\rho$), \ie the lower blob in Fig.~\ref{qhat}. 
In the subsequent section, we will evaluate the $R_{AA}$ and $v_2$ for single hadrons at two different collision energies, using both prescriptions for the transverse momentum range of the $\hat{q}$ integral. The fits to $R_{AA}$ and $v_2$ will allow us to extract the 
from of the QGP-PDF. This will depend somewhat on the prescription used. The difference will be factored into the uncertainty of the 
extracted QGP-PDF. In spite of this uncertainty, we will find that this scale dependence of $\hat{q}$ will lead to a smaller value of $\hat{q}$ at higher resolution scales, leading to an effectively smaller $\hat{q}$ at LHC, due to larger jet energies than at RHIC, at the same temperature.

\section{Single hadron $R_{AA}$ and $v_2$.}
\label{raa_v2}

In the preceding section, we outlined a model for calculating a scale and energy dependent $\hat{q}$. 
Using the principle of coherence, as demonstrated in Fig.~\ref{plot_integral}, we noted that the scale at which the medium is 
probed by a parton with a virtuality $\sim \mu^2$ is comparable to that scale. Thus as the scale of the parton increases, the effective 
$\hat{q}$ will continue to change. In fact, in the perturbative portion of $\hat{q}$, it will drop with increasing scale. In order the demonstrate this with the fewest number of parameters, we approximated $\hat{q}$ using only its perturbative component [Eq.~\eqref{first_qhat_eqn}]. This lead to the issue of dealing with the integral as $k_\perp^2$ dipped below the perturbative non-perturbative boundary $\mu_0^2$. We outlined two different prescriptions for dealing with this. In what follows, we will compute the nuclear modification factor $R_{AA}$ and azimuthal anisotropy $v_2$ of inclusive hadrons at high $p_T$, 
using both these prescriptions for $\hat{q}$. 

\begin{figure}[htbp]
\resizebox{3.1in}{3.1in}{\includegraphics{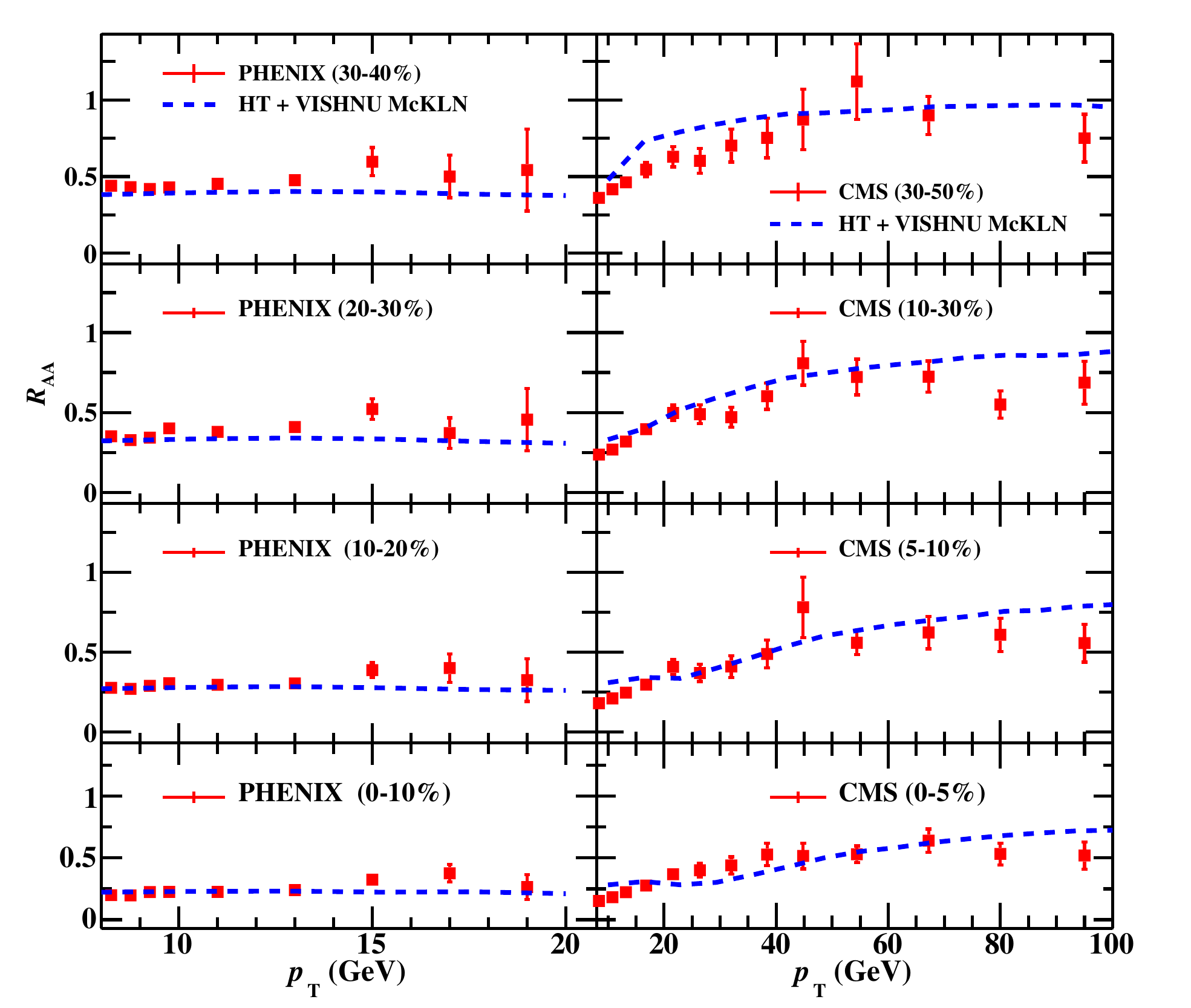}} 
    \caption{ Nuclear Modification factor at two different collision energies for four different centralities at RHIC and LHC. The parameters of the QGP-PDF are dialed to fit the bottom two panels. The scale dependence of $\hat{q}$ is calculated using prescription $[A]$ in Eq.~\eqref{A}
    and represented with the dashed lines.}
    \label{RaaA}
\end{figure}

The input to such a calculation is the form of the PDF ($x_N$ dependence) within the QGP at the lower scale of $\mu_0 = 1$~GeV. The PDF at any higher scale is 
obtained by DGLAP evolution. For this first attempt, we parametrize the PDF in standard form:
\bea
G(x,\mu^2= \textrm{1 }  { \rm GeV}^2) = N x^\alpha (1-x)^\beta.
\eea
The three coefficients are not independent of each other: The choice of $\alpha$ and 
$\beta$ restricts the choice of $N\!,$ which now replaces the overall normalization $\hat{q}_0$. 

\begin{figure}[htbp]
\resizebox{3.1in}{3.1in}{\includegraphics{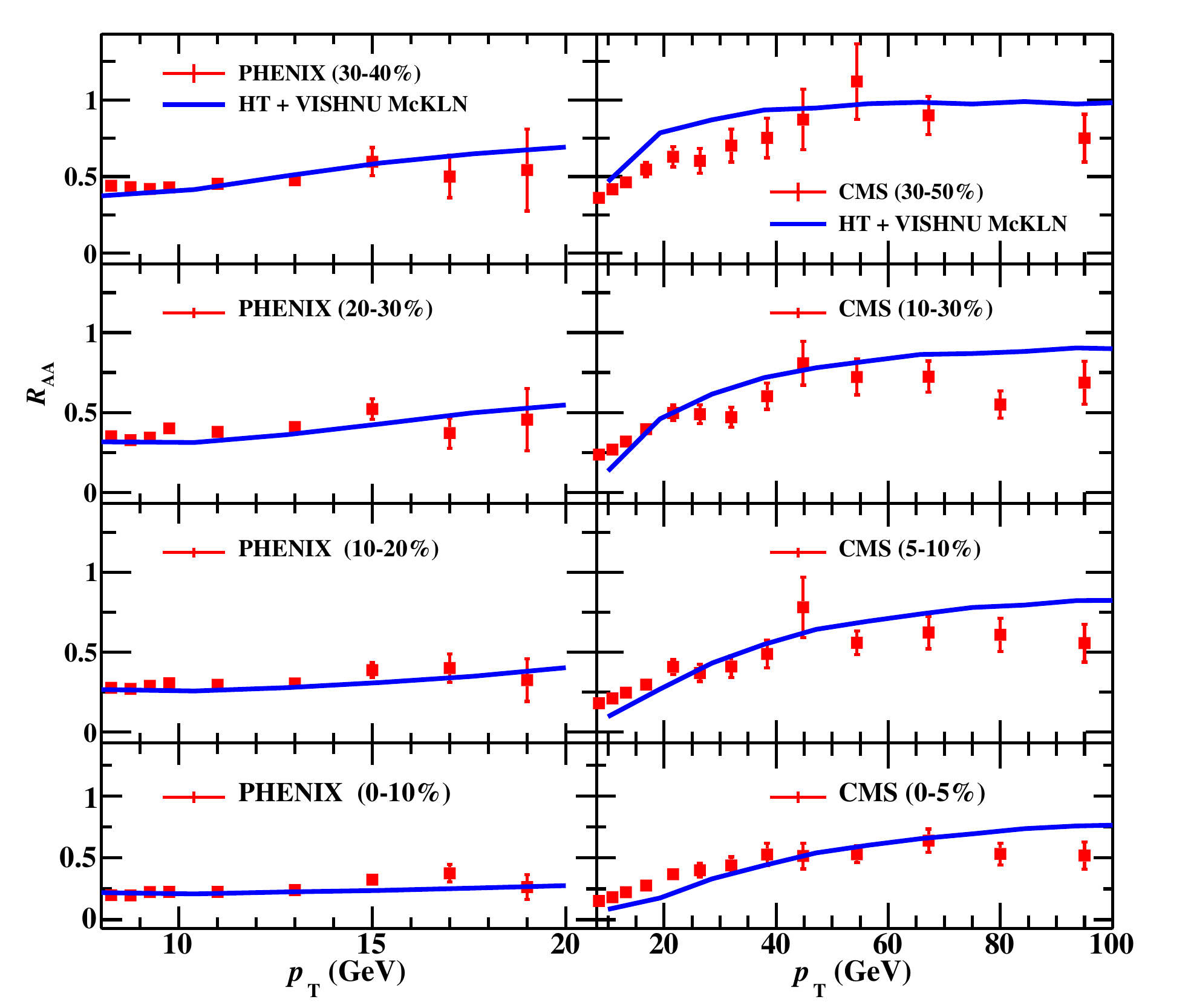}} 
%\vspace{0.25cm}
    \caption{ Same as Fig.~\ref{RaaA}, except that $\hat{q}$ is calculated using prescription $[B]$ in Eq.~\eqref{B}, and represented 
    with the solid lines.}
    \label{Raa}
%  \end{center}
\end{figure}

In this paper, we present results for the event averaged $R_{AA}$. 
Our calculations closely follow the methodology of 
Ref.~\cite{Majumder:2011uk} (See Eqs.~(2-5) in \cite{Majumder:2011uk}), which calculates the $R_{AA}$ given a $\hat{q}_0$, assuming that $\hat{q}$ scales with the local entropy density ($\hat{q} = \hat{q}_{0} s / s_{0}$, where $s_{0}=96$/fm$^{3}$). The main difference 
in the current work is the scale and energy dependence of $\hat{q}_0$, as obtained from Eq.~\eqref{qhatExpression}. 
This evolution causes a reduced $\hat{q}$ at most values of 
$x_N$ probed, leading to a natural reduction in the mean value of $\hat{q}/T^3$ at LHC compared to RHIC. 
This reduction is generic and independent of the prescription ($[A]$ or $[B]$) used to define $\hat{q}$, it is also independent of 
whether we pick a valence-like or sea-like QGP-PDF as input for the calculation of $\hat{q}$.

For the input QGP-PDF, we vary $\alpha$ and $\beta$ to obtain the combined best fit for the 0-5\% centrality bin at the LHC 
and the 0-10\% centrality bin at RHIC, as these have the 
smallest error bars. The variation with $p_T$, centrality, and now also $\sqrt{s}$ of the collision are predictions. 
The results from this particular choice of $\alpha$ and $\beta$ using prescription $[A]$ for $\hat{q}$ [Eq.~\eqref{A}], for 4 different centralities is presented in Fig.~\ref{RaaA} for both RHIC and LHC energies.
We stress once again that there is no re-normalization between RHIC and LHC energies. 
The reduction in the interaction strength $\hat{\mathscr{Q}} = \hat{q}/T^3$ is 
entirely caused by scale evolution in $\hat{q}$. The same procedure is followed for prescription $[B]$ and the results are presented in 
Fig.~\ref{Raa}. This demonstrates the somewhat mild sensitivity to the chosen prescription for the calculation of $\hat{q}$.

\begin{figure}[htbp]
\resizebox{3.1in}{3.1in}{\includegraphics{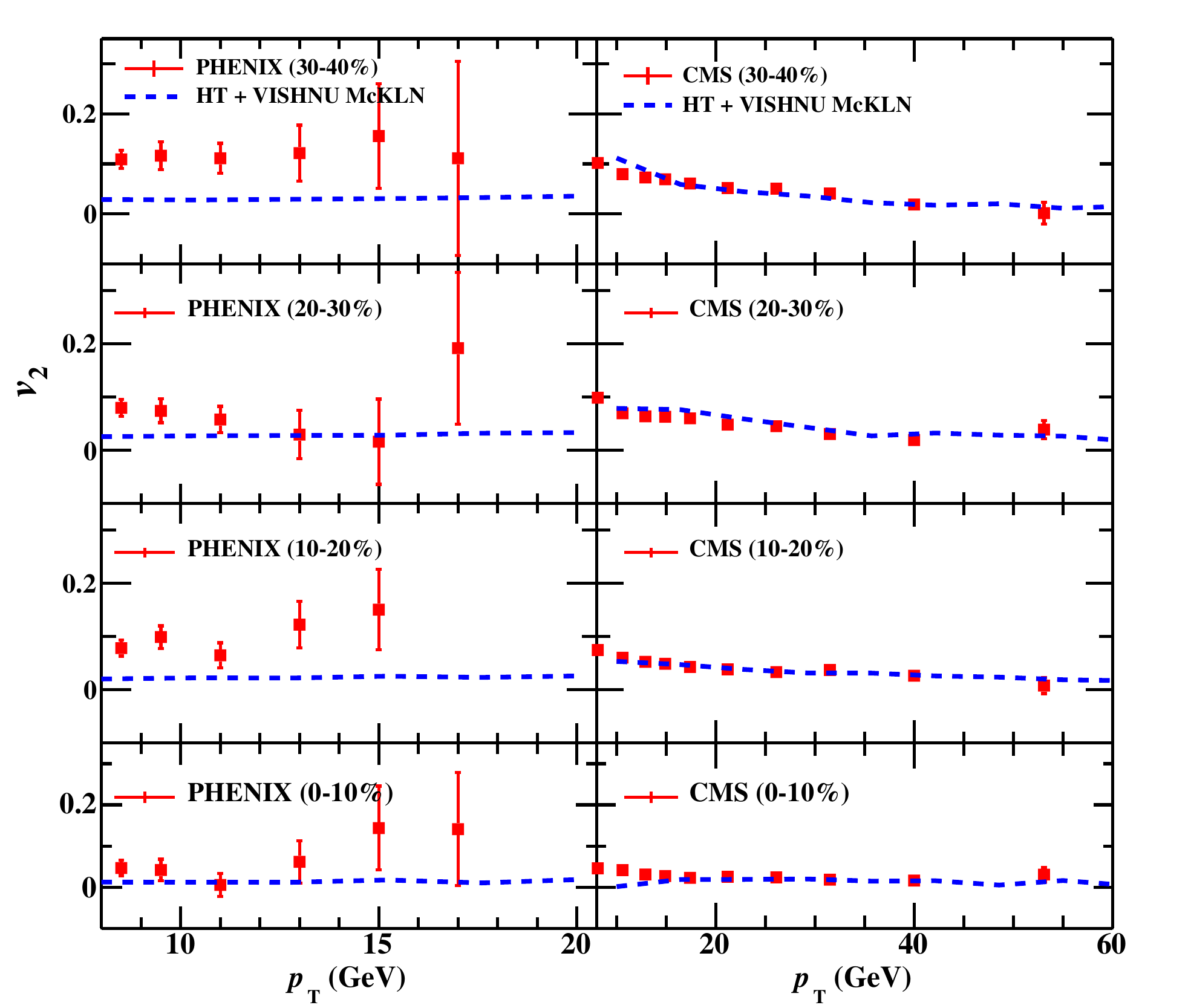}} 
%\vspace{0.25cm}
    \caption{Azimuthal anisotropy at two different collision energies for four different centralities at RHIC and LHC. Calculations are 
    carried out using prescription $[A]$ for $\hat{q}$, and represented with the dashed lines.}
    \label{v2A}
%  \end{center}
\end{figure}

We also compare the azimuthal anisotropy $v_2$ from the angle dependent $R_{AA}$, for 4 different centralities at RHIC and LHC energies in Fig.~\ref{v2}, where 
\bea
\mbox{}\!\!\!\!R_{AA} (p_T,\phi) = R_{AA} \left[  1 + 2 v_2 \cos( 2 \phi - 2 \Psi)  + \ldots \right].
\eea
where $\Psi$ is the event plane angle determined by the elliptic flow of soft hadrons. 
As the quenching of jets is carried out on a ``single-shot'' or event averaged hydro calculation, there is a well defined event plane in these fluid dynamical simulations. 
In Fig.~\ref{v2A}, we plot the azimuthal anisotropy for the case where $\hat{q}$ is calculated using prescription $[A]$. In Fig.~\ref{v2} we plot the $v_2$ using prescription $[B]$.
One should note that both Figs.~\ref{v2A},\ref{v2} represent parameter free calculations. All input parameters have been set in the angle integrated $R_{AA}$ calculations presented in Figs.~\ref{RaaA},\ref{Raa}. 

\begin{figure}[htbp]
\resizebox{3.1in}{3.1in}{\includegraphics{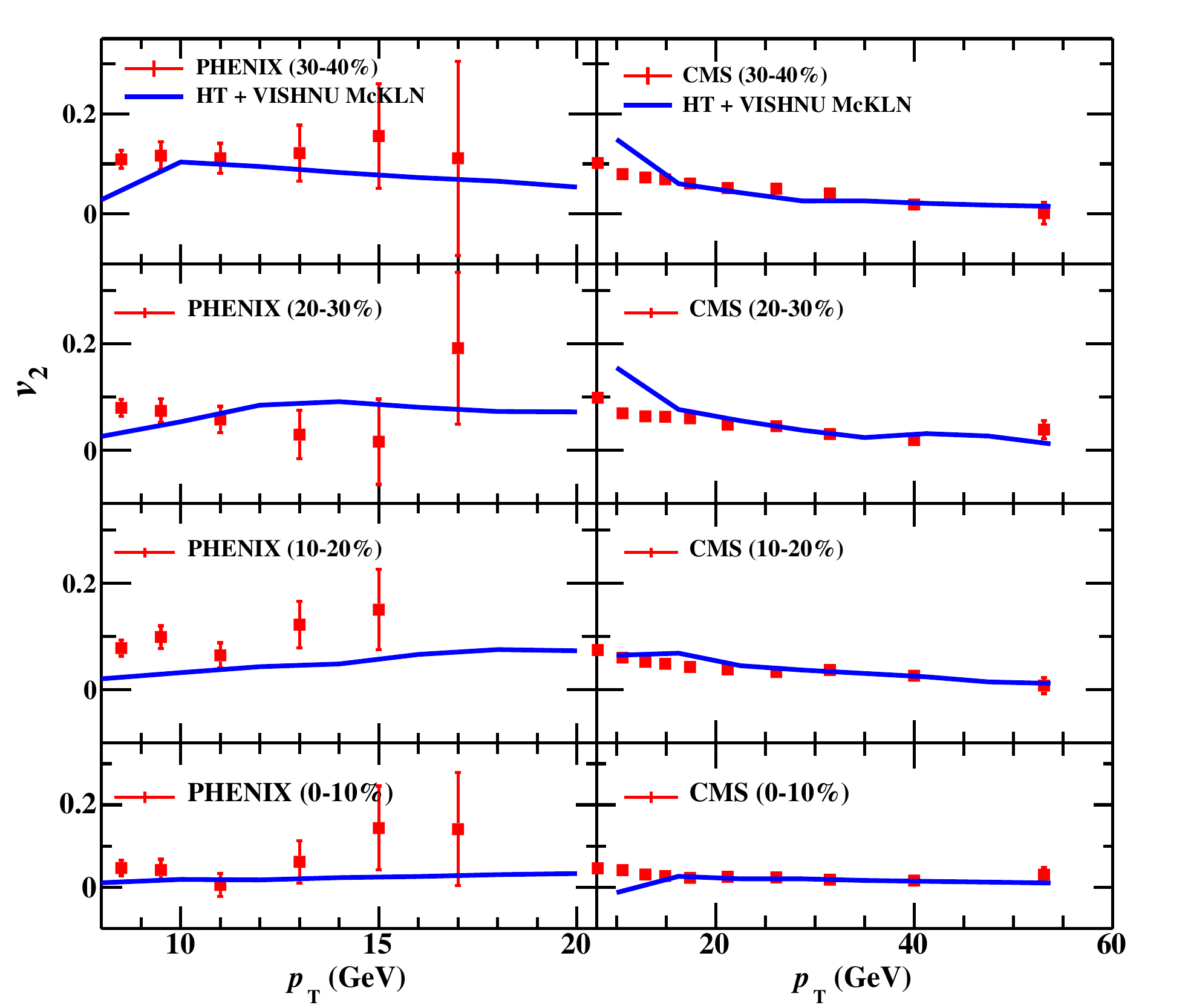}} 
%\vspace{0.25cm}
    \caption{Same as Fig.~\ref{v2A}, except calculations are carried out using prescription $[B]$ for $\hat{q}$ and represented using solid lines.}
    \label{v2}
%  \end{center}
\end{figure}

The improved fit with experimental data ($\chi^2$/d.o.f=4.8 for Fig.~\ref{RaaA} and 5.6 for Fig.~\ref{Raa}) 
without the need for an arbitrary renormalization of $\hat{\mathscr{Q}}$,  
between RHIC and LHC energies indicates that scale dependence of $\hat{q}$ 
represents an actual physical effect for jets traversing a QGP. 
This fit also adds confidence in the input PDF at $\mu_0^2 = 1$~GeV$^2$, within a QGP constituent. We have attempted several values of 
$\alpha$ and $\beta$ and 
obtained minima in $\chi^2$/d.o.f. This allowed us to isolate the input distribution to lie within the bands in Fig.~\ref{InputPDF}. 
The preferred shape does depend on the choice of prescription for the calculation of $\hat{q}$. Using prescription $[A]$ we obtain a
minimum for a sea-like PDF. This is indicated by the dashed blue line in Fig.~\ref{InputPDF}. The resulting $R_{AA}$
 and $v_2$ from this particular input PDF is also indicated by the blue dashed lines in Figs.~\ref{RaaA},\ref{v2A}. 
Using prescription $[B]$, the isolated range of input 
distributions has a ``valence'' like bump around $x_N\!\sim\!0.8$ and a large ``sea'' like contribution at small-$x_N$. The wide bump around $x_N\!\sim\!0.8$ would be consistent 
with that obtained from a temperature dependent plasma of quasi-particles. The large sea contribution likely represents the strong interaction between these.

The uncertainty bands from the input distributions for both $[A]$ and $[B]$ prescriptions overlap and are reported together as the systematic error in the extracted QGP-PDF. We should point out that while the reader may discern certain properties of the QGP by looking  
at the extracted PDF, this is still somewhat premature given the uncertainty. What is certain is that regardless of the choice of prescription of calculating $\hat{q}$ and the uncertainty in the extracted PDF, the reduction in $\hat{q}$ with increasing virtuality of the hard parton leads to a good fit to data at both RHIC and LHC energies without the need for any refitting between them. 
We have also explored the case where $M_N$ is increased to 2 GeV; this choice deteriorates the fit, increasing the $\chi^2$/d.o.f. beyond 8. This confirms our assumption that a mass of 1 GeV captures more than a degree of freedom within the  QGP.

\begin{figure}[htbp]
\resizebox{1.5in}{1.5in}{\includegraphics{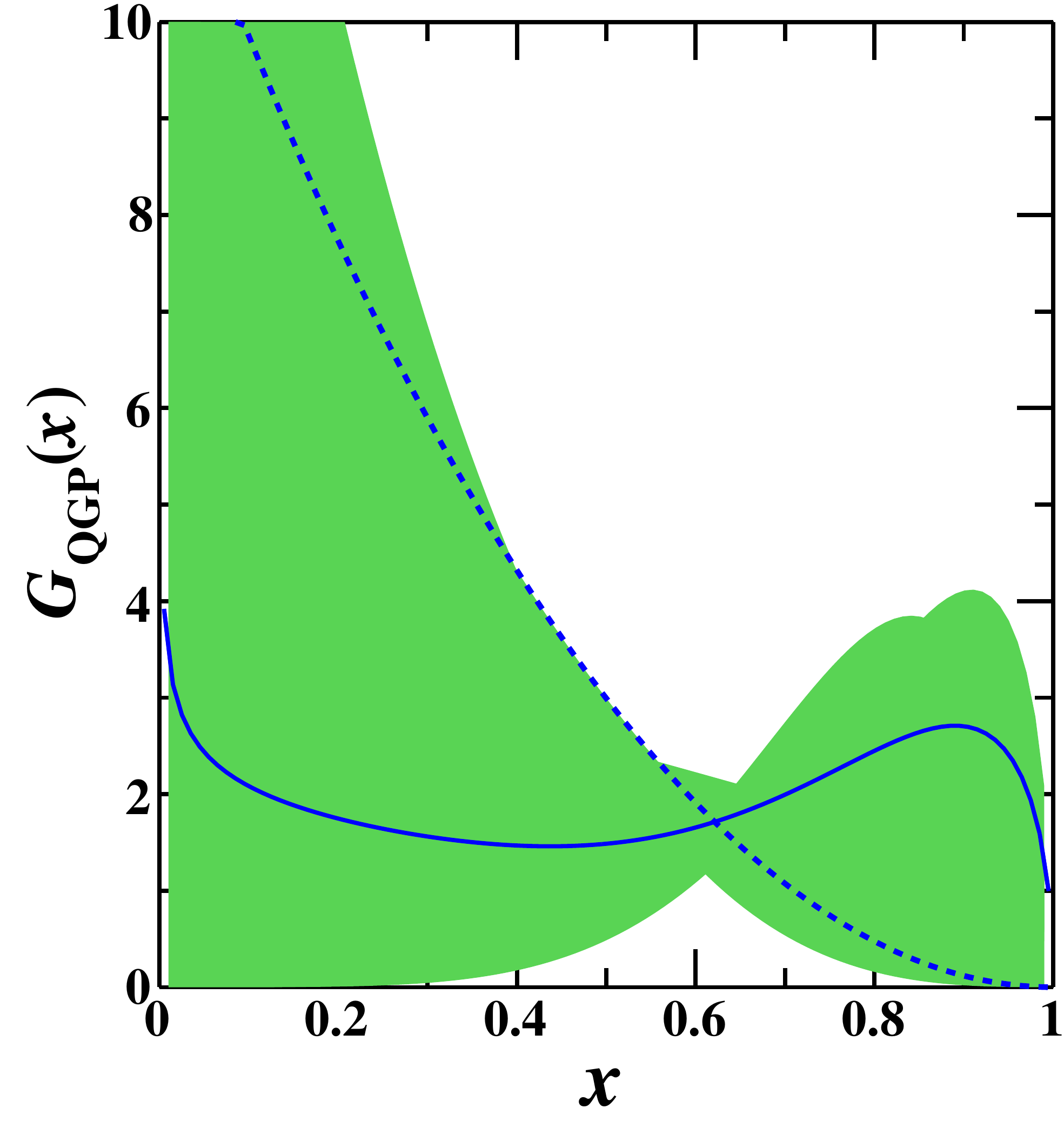}} \hspace{0.2cm}\resizebox{1.5in}{1.5in}{\includegraphics{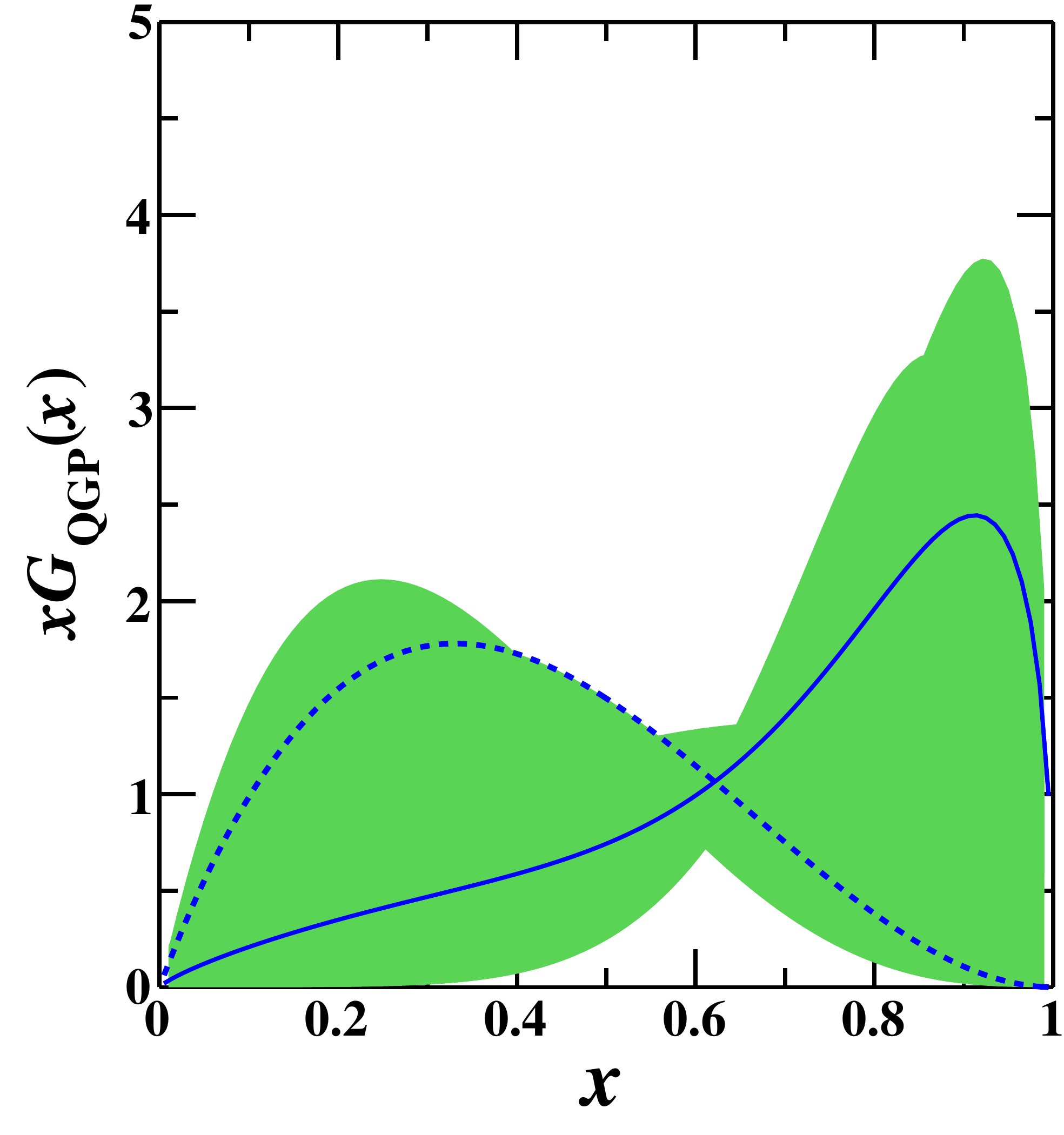}} 
%\vspace{0.25cm}
    \caption{Color Online: Left: The input PDF [$G(x_N)$] at $\mu^2=1$~GeV$^2$, used in the calculation of the scale dependence of $\hat{q}$. 
    The dashed blue line corresponds to the minimum in $\chi^2$ when $\hat{q}$ is calculated using prescription $[A]$ [Eq.~\eqref{A}],  
    The solid blue line corresponds to the minimum in $\chi^2$ when $\hat{q}$ is calculated using prescription $[B]$ [Eq.~\eqref{B}]. 
       Right: $x_N$ times the PDF. The band in both plots represents 
    the range of functions that yield very similar $\chi^{2}$/d.o.f. when compared with Fig.~\ref{Raa}.}
    \label{InputPDF}
%  \end{center}
\end{figure}

Even though the extracted PDF shows some differences between the two prescriptions for the calculation of $\hat{q}$, the resulting $\hat{q}$ as a function of scale, once the parameters have been set, are actually quite similar. In the left panel of Fig.~\ref{k_perp_int_limit} 
we show the $\hat{q}$ as a function of scale $\mu^2$, for partons with an energy of $50$~GeV for the two different prescriptions. In the case of prescription $[A]$ we used the best fit sea like distribution, for prescription $[B]$ we used a valence like distribution. For the case of prescription $[B]$, we have also shown the effect of replacing the upper limit of the exchanged momentum with its limit given by $K_\perp^2$, this may be given by either the kinematic bound ($ 2 x_N p^+ q^- $) or the scale $Q^2$, up to which the medium 
modified fragmentation function is being evolved.

Using the best fit PDF for prescription $[A]$, we also calculate the size of the $\lc k_\perp^4 \rc/\mu^2/L^-$ terms. These are terms that arise from the twist expansion and were highlighted in Eq.~\eqref{HigherOrder}. 
Typically these are ignored in all higher-twist calculations of energy loss.
In this case, since we have an explicit model of $\hat{q}$, we demonstrate that these terms are indeed quite smaller than terms such as 
$\lc k_\perp^2 \rc/L^-$ which yield $\hat{q} L /\mu^2$ and are retained in the calculation. This is an \emph{aposteriori} justification of the 
twist expansion carried out in Sect.~\ref{setup}.

\begin{figure}[htbp]
\resizebox{1.5in}{1.4in}{\includegraphics{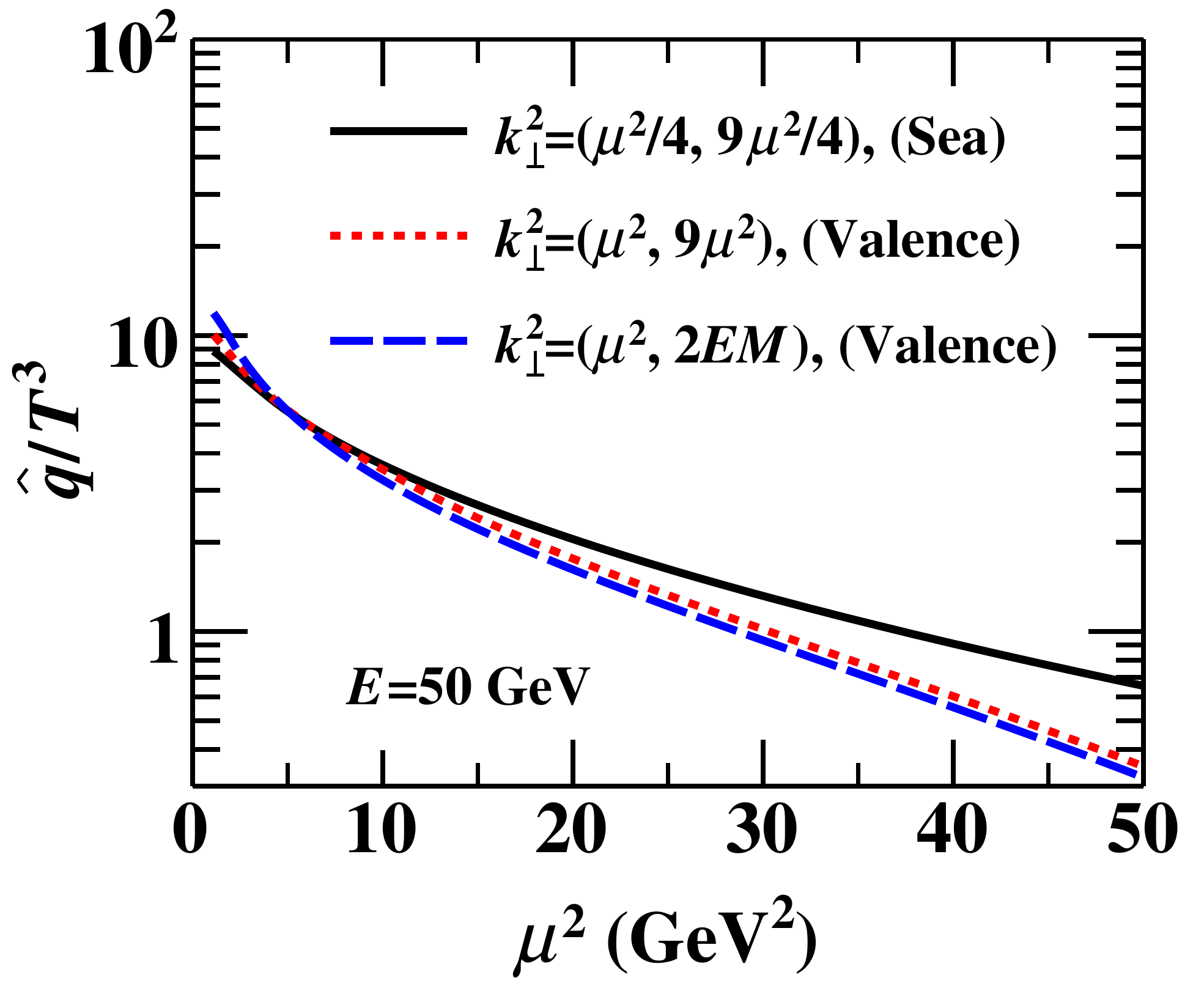}} 
\hspace{0.1cm}
\resizebox{1.5in}{1.4in}{\includegraphics{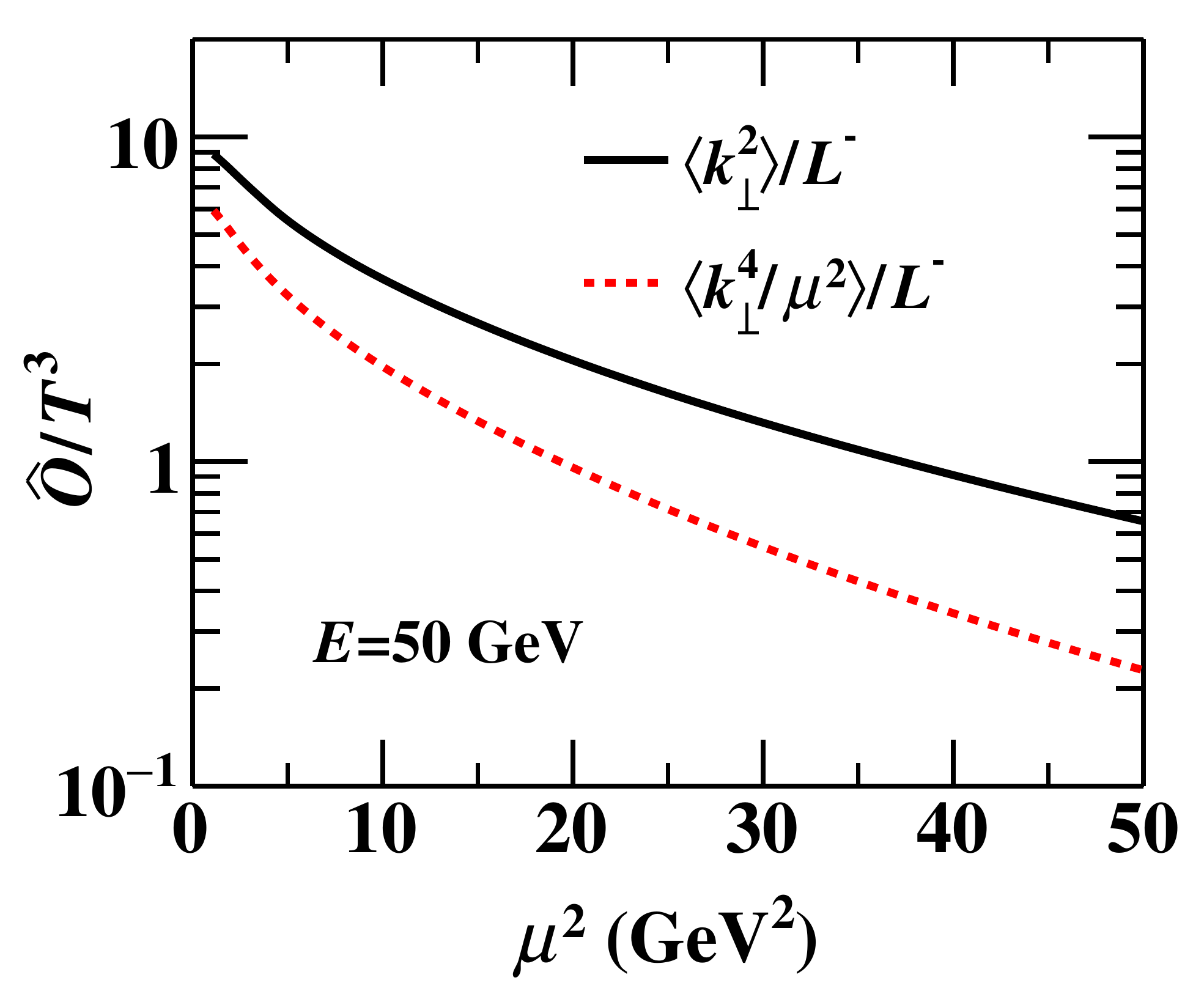}} 
%\vspace{0.25cm}
    \caption{(Color Online) Left: Calculations of $\hat{q}/T^3$ using different choices of the scale for the input PDF in Eq.~\eqref{qhat_mu_9mu}, such as the exchanged transverse momentum scale $k_\perp^2$ (black solid line), the lower limit of the exchanged momentum, 
    $\mu^2$ (red dotted line). The blue dashed line shows the minimal change obtained by changing the upper limit from $9\mu^2$ to the kinematic bound of $K_\perp^2 = 2 x p^+ q^-$ [Eq.~\eqref{qhatExpression}]. Right: A comparison between $\hat{q} = \lc k_\perp^2 \rc/L^-$  and 
   the fourth moment of the transverse momentum exchange $\lc k_\perp^4 \rc/L^-$ (scaled by the median value of $k_\perp^2 \sim \mu^2$), both scaled by $T^3$, in the model of Eq.~\eqref{qhatExpression}.  }
    \label{k_perp_int_limit}
\end{figure}

\section{Concluding Discussions.}
\label{summary}

{
In this paper, we have presented an alternative explanation to the diminishing value of the interaction strength $\hat{\mathscr{Q}} = \hat{q}/T^3$ at LHC compared to RHIC, due to the increased resolution of higher virtuality partons at higher jet energies at the LHC. As the virtuality of the partons increases with energy, the transverse size of the dipole formed by the parton and the emitted gluon decreases, and as a result, can only sample gluons from the medium that have wavelengths comparable to this size. Alternatively, the exchanged transverse momentum $k_\perp^2$ has to be of the order of the transverse momentum of the emitted gluon ($l_\perp^2$) to effect a medium induced radiation (as shown in Fig.~\ref{plot_integral}). This causes the jet parton to become sensitive to harder gluons emitted from the medium. For $k_\perp^2 > \mu_0^2 = 1$~GeV$^2$ this can be reliably calculated using a perturbative single gluon exchange formula involving a PDF of a degree of freedom within the medium. 

Several approximations were made in this first attempt involving a QGP-PDF: The calculation was formulated within the framework of single hadron production in DIS on a large nucleus. Both formulae for gluon emission for a given $\hat{q}$, and the calculation of the $\hat{q}$ using single gluon exchange with a parton from a nucleonic PDF, were then extended to the case of jets propagating through a QGP. Calculations were carried out in the 
higher-twist scheme of energy loss, where a Taylor expansion of $\lc k_\perp^2 \rc/l_\perp^2$ is carried out and only the first term is retained. This was justified in the right panel of Fig.~\ref{k_perp_int_limit} for the perturbative model used. A complete calculation of this effect would involve both a perturbative and a non-perturbative contribution to $\hat{q}$ for $k_\perp^2 < \mu_0^2$. This second contribution was ignored, to clearly highlight the diminishing behavior of the perturbative part with increasing scale of the jet parton, which was our primary goal in this first attempt. 

To numerically carry this out, we defined $\hat{q} (\mu^2)$ using two different prescriptions for the range of transverse momentum exchanged in the calculation of $\hat{q}$. This was done to restrict the lower limit of the exchanged transverse momentum to always remain in the perturbative region $\geq 1$~GeV. A comparison of these approximations for the resulting $\hat{q}$ as a function of the scale is presented in the left panel of Fig.~\ref{k_perp_int_limit}. The lack of knowledge of the mass of a QGP degree of freedom, led us to define the nucleonic $x_N$, the minimum of which is $x_N = k_\perp^2/2ME$ where $E$ is the energy of the jet parton, and $M=1$~GeV, assuming that the mass of a QGP degree of freedom will be less than 1~GeV. We have increased this to 2~GeV and found a slightly worse fit to the data. In spite of these approximations, we have clearly demonstrated that in the regime where $\hat{q}$ can be estimated perturbatively, using Eq.~\eqref{qhatExpression}, there is a clear diminishing of the interaction strength with increasing parton energy and virtuality at LHC compared to RHIC. Also, the obtained suppression in $\hat{q}$ is consistent with that required by the experimental data. In the process, we have isolated the possible PDF of a QGP degree of freedom.

In this paper, we have presented the first successful attempt to explain the JET puzzle: the downward renormalization of the interaction strength $\hat{\mathscr{Q}}$, at 
the same temperature, at LHC energies compared to RHIC energies. This was achieved by allowing $\hat{q}$ to vary with the scale of the jet. 
Fits with experimental data on $R_{AA}$ and $v_2$ at 4 different centralities from both RHIC and LHC were presented.
The inferred input distribution at the lowest possible perturbative scale of 1~GeV$^2$, has a valence-like bump and a large sea-like distribution.
The bands in Fig.~\ref{InputPDF} represent the uncertainty in the input PDF. In future efforts, we intend to carry out a more extensive study using both non-perturbative and a perturbative contribution to $\hat{q}$, involving both leading hadrons and jets. 
}
\section{Acknowledgements}
This work was supported in part by the National Science Foundation (NSF)
under grant number PHY-1207918, and within the framework of the JETSCAPE collaboration under grant number ACI-1550300, by the U.S. Department of Energy (DOE)  
under grant number DE-SC0013460.

\bibliography{q-hat-Paper_PRC}

\end{document}